\documentclass[aps,twocolumn,prd,superscriptaddress,noshowpacs,nofootinbib,noshowkeys,floatfix]{revtex4}
\usepackage{graphicx}
\usepackage{amsfonts}
\usepackage{amssymb}
\usepackage{amsmath}

\usepackage{ulem}
\usepackage{color}

\renewcommand\sout{\bgroup \color{blue} \ULdepth=-.5ex \ULset}
\begin{document}

\title{Thermodynamics of the low density excluded volume hadron gas}

\author{Krzysztof Redlich}
\affiliation{Institute of Theoretical Physics, University of Wroc\l aw, PL-50-204 Wroc\l aw, Poland}
\affiliation{EMMI, GSI Helmholtzzentrum fuer Schwerionenforschung, 64291 Darmstadt, Germany}
\affiliation{%
Department of Physics, Duke University, Durham,  North Carolina 27708, USA}
\author{Kacper Zalewski}
\affiliation{Institute of Nuclear Physics, Polish Academy of Sciences, PL-31-342 Krak\'ow, Poland}
\affiliation{%
Institute of Physics, Jagiellonian University, PL-30-059 Krak\'ow, Poland}

\begin{abstract}
{We consider thermodynamics   of the excluded  volume  particles
 at finite temperature and
chemical potential, in the
low density approximation. We assume Boltzmann statistics and study the influence of the excluded volume  on an ideal gas thermodynamics at the same temperature, pressure and  numbers of particles. We show,  that   {considering  the change of the free enthalpy  due to  the excluded volume, and using}
the Maxwell identities,  one can derive relevant   thermodynamic  functions and parameters of multi-component   {gases}. The derivation is quite general as   particles may have different sizes and shapes which  can also depend on their momenta.  Besides it's  simplicity and generality, our approach
 has the advantage of eliminating
the transcendental equations occurring in earlier
studies.
 A representative example of the excluded volume thermodynamics  is the single-component gas of hard spheres. For this case, using  the viral expansion,
the validity limits of the low-density approximation are  {also} discussed.
}

\end{abstract}
\maketitle

\section{Introduction}

A detailed analysis of secondary hadrons  produced   in heavy ion collisions (HIC) in a very broad energy range from SIS to LHC  has shown that   they are of thermal origin \cite{REV,th1,th2,th3}. Particle yields  are  excellently described as  the   Hadron Resonance Gas (HRG), an uncorrelated  gas  composed of all known hadrons and resonances  constrained by   the conservation laws \cite{REV}. The HRG is also very successful to quantify the equation of state (EoS) of the hadronic phase of  QCD at finite temperature and density, as was shown recently by comparing the EoS obtained in lattice gauge theory with the predictions of the HRG model \cite{tawfik2,lgt1,lgt2,lgt3}. These results indicate that HRG is a remarkably good  approximation to QCD thermodynamics in the confined phase, and that   HIC experiments  are indeed probing the  thermal QCD medium.

The statistical partition function for  HRG is constructed for a mixture of ideal gases, nevertheless, according to the arguments by Dashen, Ma and Bernstein \cite{Dashen} and Hagedorn \cite{Hagedorn}, it includes attractive interactions through resonance contributions.

Different extensions of the HRG model  have been proposed,  including not only the attractive,  but also  the  repulsive interactions  between hadrons. The later   are experimentally  known to be important at short distances.  Clearly, the effect of both  interactions on the thermodynamics can be introduced in a model independent way, using the S-matrix approach \cite{Dashen}. In the low density approximation the resulting expression for the partition sum of the hadron gas reduces to the Beth-Uhlenbeck form \cite{Beth}, where the  interaction contribution   is  linked  to  the experimental phase shifts \cite{Weinhold}.   However, due to far incomplete data on phase shifts,   this approach can be systematically used only in selected sectors of hadron interactions \cite{Weinhold,shift1,shift2}.

A transparent phenomenological way to account for  repulsions in a hadron gas  is to abandon  the point-like  nature of particles and introduce for them finite sizes. In this way the repulsive interactions in the hadron gas are implemented  via the  excluded volumes. Exact formulae for the thermodynamic functions of the excluded volume gas are not known, but low density approximations have been 
 developed \cite{r0,r1,r2,r3,r5,GORI,GORII,BUG,GORIII,GORIV,r4}.
The  hard core hadron resonance  gas  models were  also  successfully applied in heavy ion phenomenology \cite{REV,th1,th2} and in the interpretation of recent lattice data \cite{algt,kapusta,kapusta1,gorenstein}.

 In the following, we focus on the implementation of the repulsive interactions in a hadron gas via excluded volumes, as  introduced  in Refs. \cite{GORI,GORII}.
 {We} propose a new method, { which is based  on the Maxwell identities,}  {to calculate}  the thermodynamic functions and parameters of the low density excluded volume gas,
subject to Boltzmann statistics. {They are expressed} in terms of the excluded volumes for pairs of particles,  and the thermodynamic functions and parameters of the ideal gas at the same temperature, pressure and number of particles of each kind. The derivation is  {quite}  general, {as}  particles may have different sizes and shapes, which may also depend on their momenta.  Besides its  simplicity and generality, our approach has the advantage of eliminating the transcendental equations occurring in earlier {studies}. {The} results from {Refs.} \cite{GORI,GORII,BUG,GORIII} follow as special cases from our analysis, except for  the partial pressures in an $r$-component gas of hard spheres {presented in Ref. \cite{GORIII}}, where some additional assumptions, 
 or the  higher order  {corrections}, {are} needed.

A system  of hard spheres with all particle volumes being equal,
 is a special  {example} of the excluded volume gas. For this case we use the known results concerning the virial expansion,   to find the validity limits of the low density approximation to the thermodynamics of the excluded volume gas.
\section{Thermodynamics of the excluded volume gas }
{When modeling  thermodynamics of the extended particles, a
dimensional scale is the  size of the excluded volumes in coordinate space, which is the input in the phenomenological approach, whereas   a basic variable is the}
 configuration integral, 
 \begin{equation}\label{eq1}
  C = \int d^{3N}x e^{-\sum_{\{ij\}}^N U_{ij}(\textbf{x}_i - \textbf{x}_j)},
\end{equation}
where the summation is {taken} over all pairs of particles. { A pair-wise potential in the exponent, is introduced such that,}   $U_{ij}(\textbf{x}_i -\textbf{x}_j) = \infty$ when particles $i$ and $j$ overlap,  and $U_{ij}(\textbf{x}_i -\textbf{x}_j) = 0$ otherwise. The occurrence  of {particle} overlap may depend on more variables than written explicitly { in Eq. (\ref{eq1})}, e.g. {it can depend} on the  orientations of the volumes associated with the  particles. Then it is understood,  that {  in Eq. (\ref{eq1})}, {the integrations}   over these additional variables {are also included}.

The definition given   { in Eq. (\ref{eq1})} is very general, {as it is}  applicable for any shapes of the particles, {which are}  not necessarily the same for all  {of them}. The only requirement is,  that {the pair-wise potential}   must be define when two given particles overlap  and when they do not.

 {Introducing}  the Mayer functions,
\begin{equation}
  f_{ij}(\textbf{x}_i - \textbf{x}_j) = e^{-U_{ij}(\textbf{x}_i - \textbf{x}_j)} - 1,
\end{equation}
{in Eq.  (\ref{eq1}),}  expanding the integrand of $C$ in powers of {$ f_{ij}$}  {up to leading order},  and performing the integrations,  {one}  finds, that
\begin{equation}\label{conint}
  C = V^N\left(1 - \frac{1}{V}\sum_{\{i,j\}}v_{i,j}(T)\right),
\end{equation}
where
\begin{equation}\label{vol}
  v_{i,j}(T) = -\int d^3xf_{i,j}(x)
\end{equation}
are the excluded volumes for pairs of particles, {which in general,} may depend on temperature.

The evaluation of the integrals over Mayer functions {in Eq. (\ref{vol})}  can  be  {elementary}, like {e.g.} for the gas of hard spheres, or {may be more advanced, }  like for the gas of Lorentz contracted hard spheres \cite{BUG}. {In the following, we will not introduce    any  particular form of $f_{ij}$, and assume, that
 the integration in  Eq. (\ref{vol}) has been done.}

For {further}  applications we {consider} the expression,  $T\log C$. From Eq. (\ref{conint}), {one finds, that } {to leading order in} $v_{ij}$,
\begin{equation}\label{}
  T\log C = NT\log V - \frac{T}{V}\sum_{\{i,j\}}v_{i,j}(T).
\end{equation}
This yields, in particular, the equation of state
\begin{equation}\label{}
  \frac{pV}{NT} = \frac{V}{N}\frac{\partial\log C}{\partial V}|_{N,T} = 1 + \frac{1}{VN}\sum_{\{i,j\}}^Nv_{i,j}(T).
\end{equation}

{For the case of $N$ identical particles with   the unique two-particle excluded volume  $2v(T)$, one finds,\footnote{ For N identical particles, one needs to introduce the combinatorial factor,  ${1}/{N!}$,  into the configurational integral (\ref{eq1}).} that  in the large  $N$-limit}
\begin{equation}\label{}
  T \log C = NT\left(\log\frac{V}{N} - \frac{N}{V}v(T)+1\right),
\end{equation}
and
\begin{equation}\label{eqs1}
  \frac{pV}{NT} = \frac{V}{N}\frac{\partial\log C}{\partial V}|_{N,T} = 1 + v(T)n,
\end{equation}
where  {n=${N}/{V}$,  is the particle density}. Note, that when the particles are not identical, but all  $v_{i,j}(T)$ are {equal}, {then} only the combinatorial factor in the configuration integral changes, so that the equation of state remains {unchanged}.

In some approaches,  the excluded volume depends on the momenta of the particles  (cf. e.g. \cite{BUG}). Then,  the Mayer function for particles $i$ and $j$ depends on their momenta and it is necessary to average it over the Boltzmann distribution. As a result, the coefficients $v_{i,j}$ become temperature  dependent.  Having made this change, the terms of first order in the excluded volumes are correctly reproduced. However, since the average of the square of the excluded volume is not equal to the square of its average, this simple method works only up to first order. This is enough for the low density approximation, {however}  not for the more precise calculations needed when discussing  {its} validity limits.


Let us consider  an $r$-component gas  and denote by $N_i$ the number of particles of species $i$, with the total number of particles  $\sum_{i=1}^rN_i = N$. {Then, including the}
 combinatorial factor,  { $\prod_{i=1}^r N_i!$, in the
 configuration integral (\ref{eq1}), one finds, that in the large $N_i$-limit for all  $i=1,\ldots r$, }
\begin{eqnarray}\label{}
  T\log C &=& NT\log V - \frac{T}{2V}\sum_{i,j}N_iN_jv_{i,j}(T)\\ \nonumber
   &-& T\sum_iN_i(\log N_i-1),
\end{eqnarray}
and
\begin{equation}\label{}
 \frac{pV}{NT}  = 1 + \frac{1}{2NV}\sum_{i,j}N_iN_jv_{i,j}(T).
\end{equation}
 The above result   agrees with   {that obtained in Ref.} \cite{GORIII}. However, contrary to the earlier approaches, {there is no need}
  to solve any   transcendental equations to describe relevant thermodynamic functions.
 The reason is that we use a different set of parameters as independent variables. For the description of a state of the system any complete set of parameters is equally acceptable. E.g. for the single-component gas, one can use the parameters $(T,p,N)$ or $(T,V,\mu)$. The fact,  that in some process $N$ changes at constant $\mu$, or the other way round, is irrelevant here. {Indeed,}  {the} pressure, temperature and chemical potential are related by the {following}  {equation} \cite{GORIII},
\begin{equation}\label{gor}
  p = a(T)e^{-b(T)p + \mu/T}.
\end{equation}
In Ref. \cite{GORIII} this equation is used to find  $p(T,\mu)$, which requires the solution of a transcendental equation. The calculation of $\mu(p,T)$, from {Eq. (\ref{gor})}  is, on the other hand, elementary.

\section{{ Maxwell identities and their implications} }

Elementary phenomenological thermodynamics will now be used to find   the  thermodynamic functions and parameters of the excluded volume gas at low density, in terms {of}  the excluded volumes $v_{i,j}(T)$, and the parameters and thermodynamic functions of an ideal gas at the same temperature, pressure and numbers of particles. The parameters and thermodynamic functions of the ideal gas will be  {denoted}  with the superscript,  $id$.

Let us  {consider first}  a single-component  {gas of particles with the }   excluded volume for a pair of  particles,  $2v(T)$. In this  case, the parameter $v(T)$ may be interpreted as the single-particle excluded volume. Now we {introduce}  the excluded volume $2\lambda v(T)$. Then   {for} $\lambda = 0$, we have an ideal gas,  and for $\lambda = 1$,  the gas we are interested in. The  {point}  is to find, how the thermodynamic functions and parameters change, when $\lambda$ increases from zero,  {where} the familiar relations for an ideal gas hold, to one. The changes {are calculated} at constant temperature, pressure and number of particles.

{We consider}   {the}  differential of the free enthalpy
\begin{equation}\label{freent}
   dG = -SdT + Vdp + \mu dN + pNv(T)d\lambda,
 \end{equation}
 {where}  {the}  last term on the right-hand side is the work done on the system when the excluded volume of each particle grows. This work is done against the pressure in the surrounding of the particle, which is equal to the external pressure $p$. The low density approximation is implied by the form of this term. In the exact theory,  the excluded volumes can overlap and, therefore, the increase of the total excluded volume is slower than linear in $N$. For a low density gas, however, these overlaps can be neglected and formula (\ref{freent}) holds.

 Since free enthalpy is a well-defined function of state, the following  {Maxwell identities}  are valid
  \begin{eqnarray}\label{}
  \frac{\partial\mu}{\partial\lambda}|_{N,T,p} &=&
  \frac{\partial pNv(T)}{\partial N}|_{T,p,v} = pv(T),\\
  \frac{\partial V}{\partial \lambda}|_{N,T,p} &=&
  \frac{\partial pNv(T)}{\partial p}|_{T,N,v} = Nv(T),\\
  -\frac{\partial S}{\partial\lambda}|_{N,T,p} &=&
  \frac{\partial pNv(T)}{\partial T}|_{N,p,v} = pN\frac{dv(T)}{dT}.
\end{eqnarray}
Integrating {the above equations} at constant temperature, pressure and number of particles from $\lambda=0$ to  $\lambda = 1$,  {one gets}
\begin{eqnarray}
  \mu(T,p) &=& \mu^{id}(T,p) + pv(T)  \\
  V(T,p,N) &=& V^{id}(T,p,N) + N v(T), \\
  S(T,p,N) &=& S^{id}(T,p,N) - Np\frac{dv(T)}{dT}.
\end{eqnarray}
For the energy,  this  implies 
\begin{eqnarray}\label{energy}
  E(T,p,N) &=& 
E^{id}(T,p,N) - TNp\frac{dv(T)}{dT}.
\end{eqnarray}
Equivalent  {expressions for thermodynamic functions}   have been obtained in  {Ref. } \cite{BUG}, with  the substitution,  $NT = pV$, in the correction term {in Eq. (\ref{energy})}, which is legitimate at first order in $v(T)$.

For a Bose gas the well-known condition,  $\mu^{id}(T,p)<m$, where $m$ is the mass of a particle, implies {that}
 \begin{equation}\label{}
   \mu(T,p) < m + pv.
 \end{equation}
This upper bound is equal to the minimum free enthalpy needed to introduce one more particle into the system. It consists of the rest energy of the particle $m$ and of the work $pv(T)$, at constant temperature and pressure, necessary to create the single particle excluded volume. This observation implies,  that when $v(T)$ decreases, the  {abundance}  of particles is enhanced.

Using the expressions for the parameters $V,S,E$ of the excluded volume gas in terms of the parameters of the ideal gas  from Eqs. [17,18,19], one finds  the relations between the volume densities,
\begin{eqnarray}
  n
  &=& \frac{n^{id}}{1 + vn^{id}}  \label{density1}\\
  s
  &=&
    \frac{s^{id} -n^{id}p \frac{dv(T)}{dT}}{1 + vn^{id}} \label{density2}\\
    \epsilon
     &=& \frac{\epsilon^{id}-  n^{id}Tp\frac{dv(T)}{dT}}{1 + vn^{id}},\label{den3}
\end{eqnarray}
of particle number, entropy and energy, respectively. The arguments $(T,p,N)$,  have been skipped in these  equations.
 
 {The above relations, which link the thermodynamics of the  point-like and  the excluded volume  particle gases, are consistent   with previous
studies in Ref. \cite{GORI}, and are also very successful in the applications to  thermal description of particle production yields  in HIC \cite{REV,th1,th2}.}

The { method introduced above,  for the  single-component gas,  can be generalized to  multi-component system. } For an  $r$-component gas the differential of the free enthalpy   {reads},
\begin{eqnarray}\label{ren}
   dG &=& -SdT + Vdp + \sum_i^r\mu_i dN_i \\ \nonumber
   &+& \frac{1}{2}\sum_{ij} {p_iN_j}{}v_{i,j}(T)d\lambda,
 \end{eqnarray}
{where we have introduced, $
  p_i = \frac{N_i}{N}p$, the relation which is valid  for ideal gases, and is also   legitimate, in the
correction term {due to excluded volumes}.}

 {The} {last term in Eq. (\ref{ren}),  which describes the} work done when the excluded volumes increase,  can be interpreted as the sum of works done around all particles. The work done around a particle of type $i$ is a sum of $r$ terms: the work needed to {free} the volume $v_{i,1}(T)d\lambda$ from particles of type one against the partial pressure $p_1$, the work needed to free the volume $v_{i,2}(T)d\lambda$ from particles of type two against the partial pressure $p_2$, and so on,  up to particles of type $r$. Thus, for the multi-component gas the one-particle excluded volume can be defined, {but} it does not have a simple geometric {meaning}.

{Moreover,}  {there} is some ambiguity in dividing the work associated with the increase of the excluded volumes among the surroundings of the particles. Clearly, the term
${p_1N_2}v_{12}d\lambda${=$p_2N_1v_{12}d\lambda$},
contributes to the work done in the surroundings of the particles {of types} $1$ and $2$.  {Following} the interpretation presented above, for each particle of type $1$ this work is  $\frac{p_2}{2}v_{12}d\lambda$,  and for each particle of type $2$, {{it is}   $\frac{p_1}{2}v_{12}d\lambda$. According to Ref. \cite{GORIII}, however, for $r=2$ and  {for } the second kind of particles   being point-like, the contribution of this term to the work done around a particle of type $1$ is zero,  and around a particle of type $2$ {it is}  { ${p_1}{}v_{12}(T)d\lambda$}. Both {these} interpretations are consistent with our { leading order expression}  for the differential of free enthalpy.

{For an $r-$component gas, the} Maxwell identities are {generalized as}
  \begin{eqnarray}\label{}
  \frac{\partial \mu_i}{\partial \lambda}|_{N,T,p} &=&\frac{\partial \left(\frac{1}{2}p\sum_{i,j}\frac{N_iN_j}{N}v_{i,j}\right)}{\partial N_i}|_{T,p,\lambda}\\ \nonumber
 &=&  p\sum_j\frac{N_j}{N}v_{i,j} - \frac{1}{2}p\sum_{i,j}\frac{N_iN_j}{N^2}v_{i,j},\\
  \frac{\partial V}{\partial \lambda}|_{N,T,p}\nonumber
   &=& \frac{\partial \left(\frac{1}{2}p\sum_{i,j}\frac{N_iN_j}{N}v_{i,j}(T)\right)}{\partial p}|_{N,T,\lambda}\\
&=&\frac{1}{2}\sum_{i,j}\frac{N_iN_j}{N}v_{i,j}(T),\\
  -\frac{\partial S}{\partial \lambda}|_{N,T,p} &=& \frac{\partial \left(\frac{1}{2}p\sum_{i,j}\frac{N_iN_j}{N}v_{i,j}(T)\right)}{\partial T}|_{N,p,\lambda}\nonumber
\\&=&\frac{1}{2}\sum_{i,j}\frac{N_iN_j}{N}\frac{dv_{i,j}(T)}{dT},
\end{eqnarray}
where the subscript $N$ indicates that all the multiplicities $N_1,\ldots,N_r$ are fixed.
Integrating {the above equations} over $\lambda$ from zero to one,  at fixed temperature, pressure and numbers  of particles, {one finds}
\begin{eqnarray}
  \mu &=& \mu^{id} + p\sum_j\frac{N_j}{N}v_{i,j} - \frac{1}{2}p\sum_{i,j}\frac{N_iN_j}{N^2}v_{i,j},  \\
  V &=&  V^{id} + \frac{1}{2}p\sum_{i,j}\frac{N_iN_j}{N}v_{i,j},\\
  S &=& S^{id} - \frac{1}{2}p\sum_{i,j}\frac{N_iN_j}{N}\frac{dv_{i,j}}{dT},
\end{eqnarray}
where the arguments in each case are $(T,p,N_1,\ldots,N_r)$.
Following, the discussion  of the  single-component gas, the generalization of Eqs. (\ref{density1}),  (\ref{density2}) and (\ref{den3}),
to the $r-$compannet gas,  is rather {transparent}.

\begin{figure*}[ht]
\centering
\vskip -1.0cm\includegraphics[width=3.1in,angle=-90]{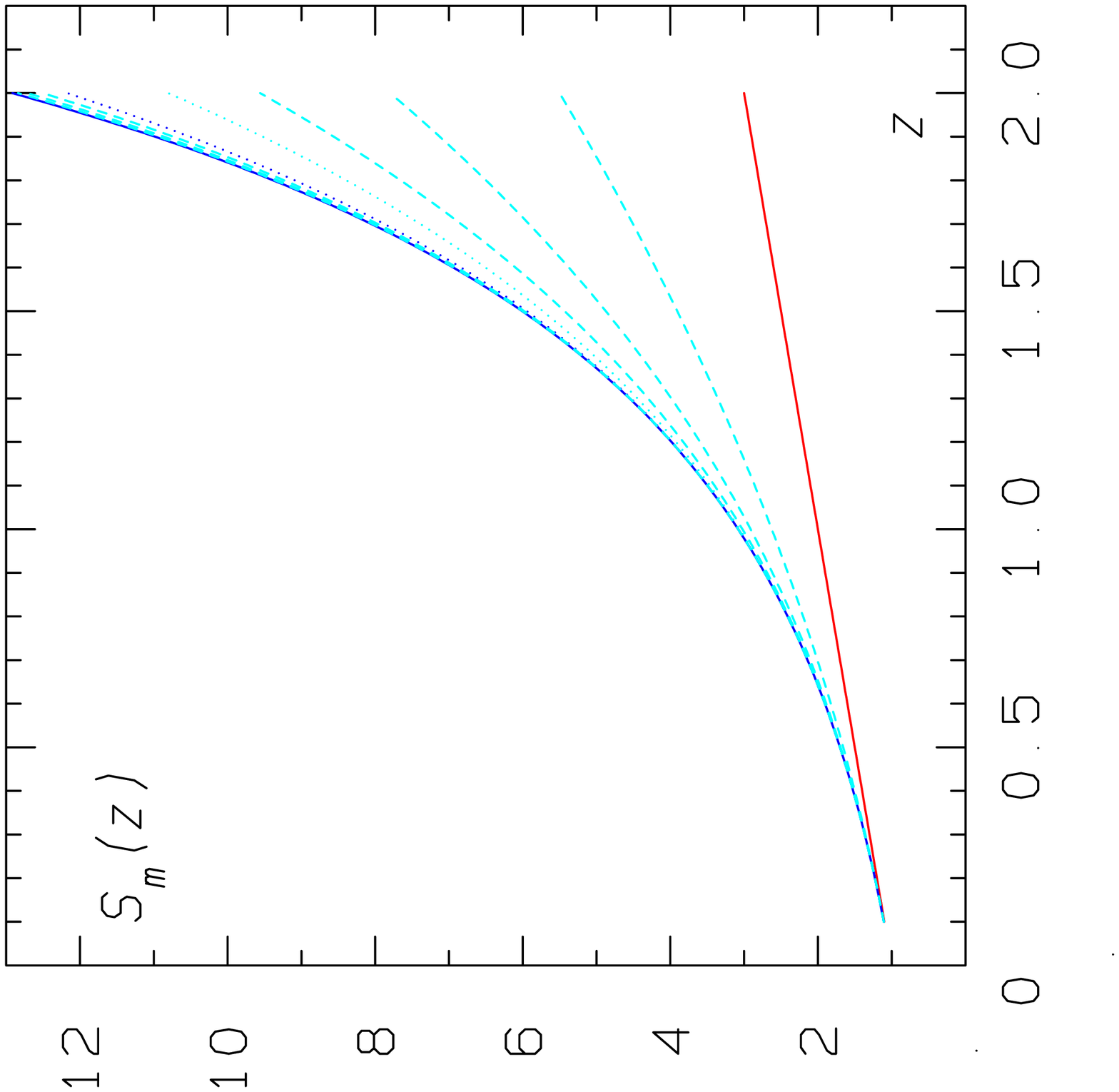}\hskip 3.6cm
\includegraphics[width=3.2in,angle=-90]{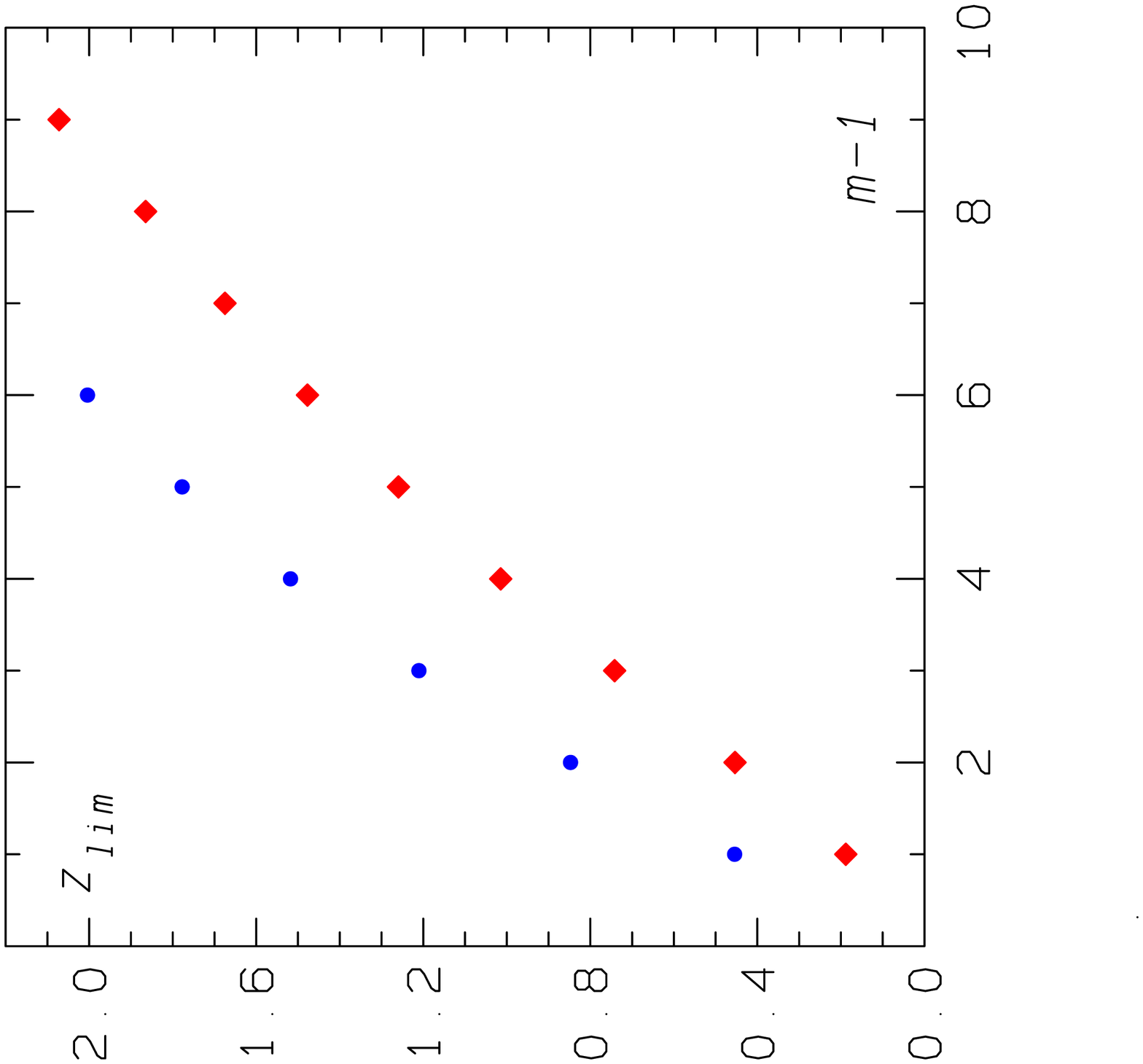}
\vskip -1.5cm
\caption{Left-hand figure: The virial series from Eq. (\ref{partial}) truncated at $m=2,\ldots,12$. The term $m=2$ corresponds to the lowest line. Right-hand figure: The limiting values $z_{lim}$ of $z$ for which the accuracy of $S_m(z)$ from Eq. (\ref{partial}) with m=2,..,10,  relative to $S_{12}(z)$, is 2\% (diamonds) and 10\% (dotes).       }
\label{fig1}
\end{figure*}

\section{Validity  of the low density approximation}
{In the previous sections,  we have derived the thermodynamic observables of a   gas
of extended particles,   under the  low density approximation.}
In order to determine reliably the validity limits of {this}  approximation,  it is necessary to compare {our results}   with the exact solution, or with   {its} good approximation. In the excluded volume gas problem,  {such comparison}   is possible   for the gas of hard spheres  {of}  the same radius $R$,
 i.e. with the volume
\begin{equation}\label{eq32}
  v_0 = \frac{4}{3}\pi R^3.
\end{equation}
The corresponding single-particle excluded volume is {then},  $v = 4v_0$. For $R=0$,  the gas reduces to the ideal gas,  and the low density approximation yields the exact result. With increasing $R$ the approximation deteriorates. {The} task is to find, up to what values of $R$,  {the low density approximation} is reliable,  within a given error margin.

 {The} equation of state of the gas of hard spheres can be written in the form of a virial expansion,
\begin{equation}\label{virser}
  \frac{p}{nT} = \sum_{i=1}^\infty b_i z^{i-1};\qquad z = nv.
\end{equation}
The virial coefficients $b_i$ are known {explicitly}  \cite{CMC,WOO} for $i=1,\ldots, 12$, {and they} are quoted in  Table 1.

\begin{center}{
 $\begin{array}{|c|r||c|r||c|r||c|r|}
 \hline
 b_1 & 1 & b_4 & 0.286949 & b_7 & 0.0130235 & b_{10} & 0.0004035
   \\ \hline
 b_2 & 1 & b_5 & 0.11025 & b_8 & 0.004183 & b_{11} & 0.000123 \\ \hline
 b_3 & 0.625 & b_6 & 0.0388819 & b_9 & 0.001309 & b_{12} & 0.000037 \\ \hline
 \end{array} $\\}\end{center}
\begin{center}
            Table 1. Virial coefficients for the gas of hard spheres from Refs. \cite{CMC,WOO}.
\end{center}

Comparing Eqs. (\ref{eqs1}) and (\ref{virser}),  {it is clear,} that the low density approximation is obtained when only the first two terms of the virial expansion are taken into account. The convergence of the virial expansion deteriorates when the parameter $z$ increases. In order to find how good the convergence is, we must define the relevant range of $z$.

Obviously, the total volume of the spheres $Nv_0$ must be smaller than the total volume of the gas. This implies, {that}
 \begin{equation}\label{}
   z < 4.
 \end{equation}
Moreover, it is not possible to pack hard spheres into a volume without leaving free spaces. {Consequently}  {for}  the closest packing
\begin{equation}\label{}
  nv_0 = \frac{\pi}{3\sqrt{2}}.
\end{equation}
An equivalent formula was conjectured by Kepler in 1611, then Gauss in 1831  demonstrated,  that this is the closest packing possible,  if the spheres form a lattice. {The general proof is difficult, however can be handled by using computers.
}
  Its completion was announced by T.C. Hales in 2014. From this limit, one gets
\begin{equation}\label{}
  z < 2.96.
\end{equation}
Finally, it is plausible,  that the hadrons form a fluid rather than a condensed phase. The condition for the freezing of the gas of hard spheres is, $nv_0 = 0.494 $ \cite{CMC}, which yields the limit,
\begin{equation}\label{zlimit}
  z < 1.976.
\end{equation}
Above this { limit},  the virial expansion for the gaseous phase   of hard spheres {system}  diverges.
{Consequently, one concludes, }  that  { the relevant}  region {of $z$, is as follows}
\begin{equation}\label{zmax}
  0 \leq z <  2.
\end{equation}
{We introduce the partial sum,}
\begin{equation}\label{partial}
  S_m(z) = \sum_{i=1}^m b_iz^{i-1},
\end{equation}
{for} $m \geq 2$. {Knowing the values of virial coefficients, one  can  quantify, how good is the  low density approximation for different values of $0 \leq z \leq 2$.}
\begin{figure*}[ht]
\center
{
\vskip -1.0cm\includegraphics[width=3.23in,angle=-90]{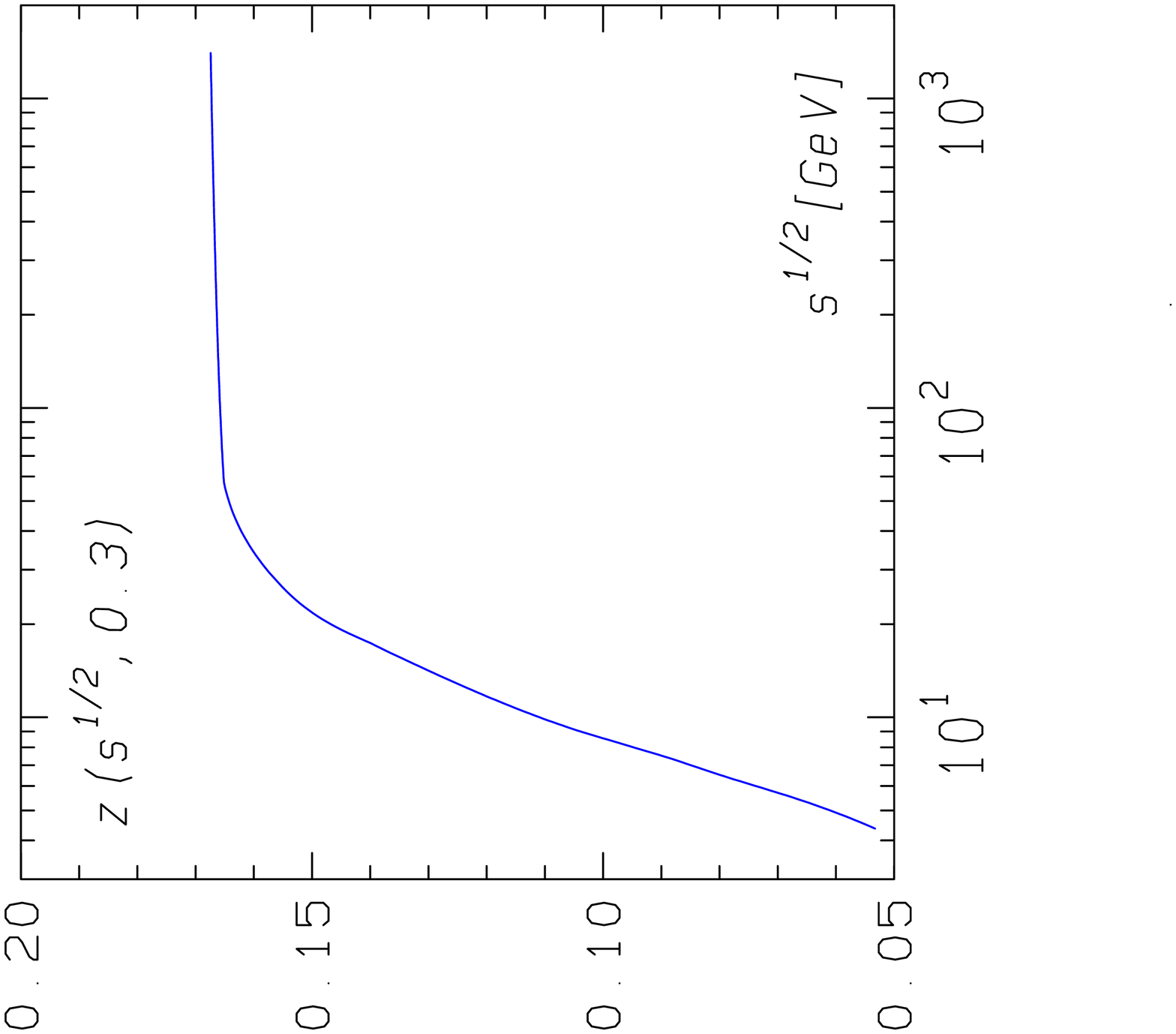}\hskip 3.6cm
\includegraphics[width=3.17in,angle=-90]{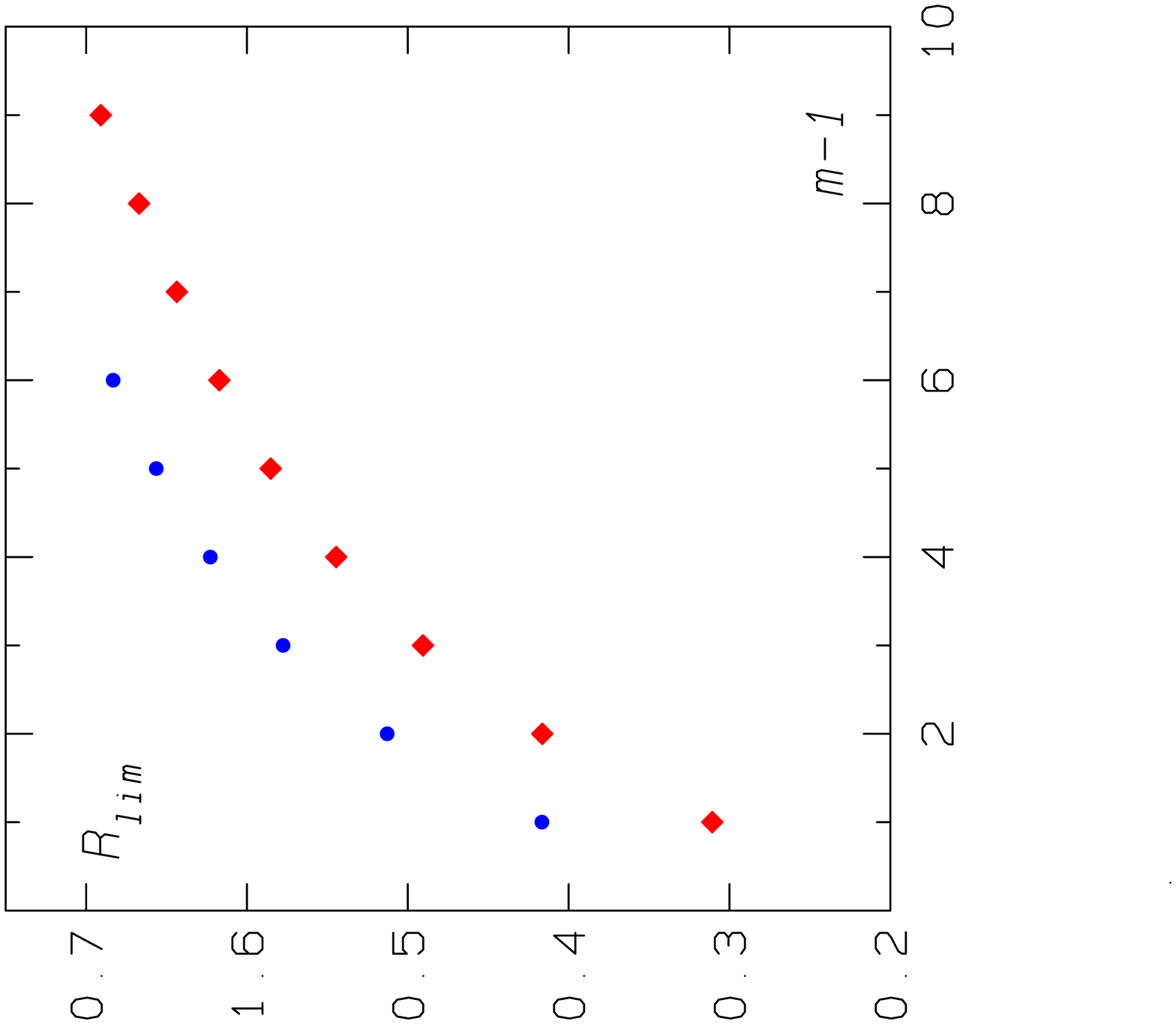}}
 \vskip -1.9cm
\caption{Left-hand figure: Function $z(\sqrt{s},0.3)$ from Eq. (\ref{eqz})  along the chemical freezout line   extracted from particle yields data  taken  in central nucleus-nucleus collisions at different collision energies \cite{REV,1gev1}.
Right-hand figure:
The limiting values of $R$ from Eq. (\ref{eqz}) at $n = 0.375$fm${}^{-3}$,   for which the accuracy of the truncated series $S_m(z)$ for  m=2,..,10, relative to $S_{12}(z)$,  is 2\% (diamonds) and $10$\% (dots).
}\label{fig2}
 \end{figure*}


{In Fig. \ref{fig1}-left, { we show}  the sum $S_m(z)$ for  $0 \leq z \leq 2$, and for different  $m = 2,\ldots,12$. The low density approximation is seen in this figure to describe the equation of state  within less than  few percent,   for $z<0.5$. With increasing $z$ up to unity, the series is known to converge  and the deviations are up to 30$\%$, whereas for $z>1$ a large corrections to the low density approximation are to be expected. For $z$ close to two, the virial expansion in Eq. (\ref{virser}) diverges.}

In Fig. \ref{fig1}-right,  we show for $m = 2,\ldots 9$ the limiting values of $z$ for which the accuracy of approximation of { $S_{12}(z)$ series by}  $S_m(z)$ is  {two}  and   ten  percent{, respectively}.   Since only the values {of} $z<2$ are of interest, the limits significantly exceeding   { $z=2$},   are not shown {in this figure}.


The Hadron Resonance Gas (HRG) formulated with excluded volume from Eq. (\ref{eq32}), was applied  to describe  particle production in heavy ion collisions in a very broad energy range from SIS up to LHC  \cite{REV}. All particle yields in heavy ion collisions were successfully quantified by the model along a common  freezeout line in the temperature and baryon-chemical potential plane \cite{REV,1gev}. This finding allows to quantify the conditions for the applicability
of the low density approximation in the HRG  of excluded volume particles, for the phenomenologically relevant system,  along the freezeout line.

In  applications, {the total density of particles at chemical freezeout is temperature and chemical potential dependent,    and consequently,
changes with the collision energy $\sqrt s$. Thus,}     $z$ in Eq. (\ref{virser})  is a function of  $\sqrt s$,  and of the chosen value of $R$. Let us  introduce  the function
\begin{equation}\label{eqz}
  z(\sqrt{s},R) = \frac{16\pi}{3}n(\sqrt{s})R^3,
\end{equation}
where $R$ is expressed in fermis.

In Fig.~\ref{fig2}-left,   we show the
energy dependence of $z(\sqrt{s} ,0.3)$
for the HRG,  which includes all  particles and resonances,  listed by
the Particle data Group.
 For simplicity,  the radius $R$ in  Eq. (\ref{eqz}) was assumed to be common for all particles and  antiparticles with the  value,  $R\simeq 0.3$ fm.

The function $z(\sqrt{s},0.3)$  is small at low energies bellow AGS, increases monotonically with increasing energy,  {and at high energies beyond SPS,  it}  saturates at the  value slightly below 0.17. In our further estimates we will use the approximation $z(\sqrt{s},0.3) = 0.17$, or equivalently,  $n=0.375$ fm$^{-3}$.

{ The upper bound of $z$, established in Eq.}     (\ref{zlimit}) {implies, in general,  the restriction on the radius of the hard core repulsion, }
 \begin{equation}\label{Rlimit}
   R < 0.680 ~\mbox{fm},
 \end{equation}
for   density   $n=0.375$ fm$^{-3}$.
One could go beyond this limit, speculating that the virial series is asymptotic, in the sense that the sum of the first terms gives a reasonable approximation,  in spite of the divergence of the series, but this is  rather risky.

In Fig. \ref{fig2}-right,  we show, for $m=2,\ldots,9$, the limiting values of $R$ for which the accuracy of the truncated series is {two}  and  ten  percent{, respectively}. For $R=0.3$ fm,  the low density approximation is good enough to get  {two}  percent accuracy. For $R=0.5$ fm,  in order to get this accuracy,  one would need to  include two more terms  {in}  the virial expansion, while for {ten}  percent accuracy, still  one more term is {needed}.

{To further improve the equation of state of a gas with  hard core repulsion, it was proposed \cite{GORI},  to consider the equation of state in the following form }
\begin{equation}\label{}
  \frac{p}{nT} = \frac{1}{1 - nv(T)}.
\end{equation}
This is equivalent {to}  putting $b_k = 1$ for all $k$, instead of neglecting them for $k>2$,  in the virial expansion (\ref{virser}). Since $b_3 = 0.625$, the error in the $O(z^2)$ term is reduced by almost a factor of two. The errors on the higher order terms are greatly increased, but for sufficiently small values of $z$ this is unimportant. Comparing with the partial sum $S_{12}(z)$,  one finds,  that the improved low density approximation is better than the standard one for $z < 0.312$, i.e. for $R<0.37$ fm.

\section{Summary and Conclusions}

{We have  discussed  thermodynamics   of the excluded  volume  particles
 at finite temperature and
chemical potential, in the
low density approximation. Assuming  Boltzmann statistics,    the influence of the excluded volume  on an ideal gas thermodynamics  has been
{derived from the change of the free enthalpy and  by using the
 Maxwell identities.   }

The calculation of the {relevant}  thermodynamic functions and parameters  of the  excluded volume gas,  in terms of the two-particle excluded volumes and of the thermodynamic functions and parameters of the ideal gas,  splits into two steps:

{ At the first step, it requires  calculating }    the two-particle excluded volumes, which is  { rather standard,} but one should keep in mind,  that the conversion of the momentum dependence of the two-particle excluded volumes into their temperature dependence,   {works}  only in the low density approximation and under  Boltzmann  statistics.
 The single particle exclusive volume is a natural concept for the single-component gas. For  {multi-}component {systems},  analogous  {excluded} volumes appear, but there are ambiguities in their definitions and they have no simple geometrical meaning.

The thermodynamical calculation {proposed here}, which is the second step, {is} valid   in the low density {approximation}. In order to {go beyond  {this limit,}    it would be necessary to correct for possible overlaps of excluded volumes.

The physical picture behind the thermodynamic calculation, introduced  here, is {based on the observation, that by increasing excluded volume $v(T)$  of particles, the} work done on the system at constant temperature (T), pressure (p) and particles number (N),  equals the increase of {the} free enthalpy. Thus e.g.,   for the single-component gas,  the free enthalpy stored in the excluded volumes is $Npv$. Its differential,
yields the corrections to the entropy, volume and chemical potential of the gas, {and through  Maxwell identities, {provides} the relation between thermodynamics of the extended and point-like particle gases. This new approach, is also valid for  the multi-component  {systems.}

A problem of  {phenomenological}  importance is that of the validity limits for the low density approximation in heavy ion collisions. Assuming that all  particles are hard spheres with the same radius $R$, we {have}  {shown, based on the virial expansion of the equation of state,} that for $R=0.3$ fm   the approximation, for the relevant range of parameters, is very good. For $R = 0.5$ fm it introduces a significant error,  and for $R= 1$ fm,  it is doubtful.

\section*{ Acknowledgments}
The authors thank Andrzej Bia\l as for fruitful discussions. One of the authors (KZ) was partly supported by the Polish National Science Center (NCN), under  grant DEC-2013/09/B/ST2/00497. K.R. acknowledges  support of  the Polish Science Center (NCN) under Maestro grant DEC-2013/10/A/ST2/00106, and of the U.S. Department of Energy under Grant No.  DE-FG02- 05ER41367. K.R. also acknowledges
fruitful discussions  with Peter Braun-Munzinger,  Steffen Bass, Bengt  Friman and Pok Man  Lo.

\end{document}